\documentclass[pre,twocolumn,floatfix,superscriptaddress]{revtex4}
\usepackage{graphicx,amsfonts,amssymb,amsmath, hyperref}
\usepackage[applemac]{inputenc}

\newif\ifhyper
\hypertrue
\ifhyper
\hypersetup{
  citecolor = {green},
  colorlinks = {true}, 
  urlcolor = {blue} 
}
\fi

\newlength{\ldag}
\begin{document}

\title{Integration through transients approach to the $\mu(\mathcal{I})$ rheology}

\author{O. Coquand} 
\email{oliver.coquand@dlr.de}
\affiliation{Institut f\"ur Materialphysik im Weltraum, Deutsches Zentrum f\"ur Luft- und Raumfahrt (DLR), 51170 K\"oln, Germany}

\author{M. Sperl} 
\email{matthias.sperl@dlr.de}
\affiliation{Institut f\"ur Materialphysik im Weltraum, Deutsches Zentrum f\"ur Luft- und Raumfahrt (DLR), 51170 K\"oln, Germany}
\affiliation{Institut f\"ur Theoretische Physik, Universit\"at zu K\"oln, 50937 K\"oln, Germany}

\author{W. T. Kranz} 
\email{kranz@thp.uni-koeln.de}
\affiliation{Institut f\"ur Theoretische Physik, Universit\"at zu K\"oln, 50937 K\"oln, Germany}
\affiliation{Institut f\"ur Materialphysik im Weltraum, Deutsches Zentrum f\"ur Luft- und Raumfahrt (DLR), 51170 K\"oln, Germany}


\begin{abstract}
	This work generalises the granular integration through transients formalism introduced by Kranz \textit{et al.}
	[Phys. Rev. Lett. \textbf{121}, 148002 (2018)] to the determination of the pressure.
	We focus on the Bagnold regime, and provide theoretical support to the empirical $\mu(\mathcal{I})$ rheology laws,
	that have been successfully applied in many granular flow problems.
	In particular, we confirm that the interparticle friction is irrelevant in the regime where the $\mu(\mathcal{I})$ laws
	apply.
\end{abstract}

\maketitle

\section{Introduction}

	Granular matter can exist in a variety of states, including the usual solid, liquid and gas \cite{Jaeger96,
	Aranson06,Andreotti13}.
	Indeed, even in the simplest dry granular systems where no attractive force is present, the phenomenology of moderately high density flows
	that present both significant correlations between particles and frequent collisions is significantly different from that of the low-density
	gaseous phase where kinetic theory can be applied \cite{Andreotti13}.
	Up to now, no unified theory of liquid granular flows has been shown to be fully successful in describing their rich phenomenology.

	One of the greatest challenges of any granular liquid theory is to be able to account for its fluid characteristics, as well as its
	ability to become progressively solid-like as its density increases.
	In the first attempts to build models of granular liquid flows, this was overcome in the following way \cite{Savage79,Savage89,Savage98,Ancey99}:
	The fluid behavior was modeled through a Navier-Stokes-like equation of momentum conservation, and yielding was accounted for by coupling
	this equation to a Mohr-Coulomb criterion similar to the one used in soil mechanics:
	The material is described through the use of a characteristic quantity $\mu$, called the effective friction, and that plays a role similar to that of
	the friction coefficient in Coulomb's laws of solid friction.
	Whenever the tangential stress $\sigma_0$ applied to the material is greater than $\mu$ times the normal
	stress,	the material begins to flow.
	Despite its simplicity, this model describes many of the properties of dense granular flows, and is still widely used in geophysics \cite{Kelfoun08,Kelfoun09,Frey10,Ogburn17,
	Salmanidou17,Delannay17,Gueugneau17,Gueugneau19,Pahtz20}.

	Then, a natural way to proceed in fluid mechanics is to identify the relevant flow regimes and characterize
	them with dimensionless numbers.
	Already in the very first studies, the importance of a number characterizing the competition between fluidization
	and collisional stresses had been identified \cite{Savage89,Ancey99}.
	This number is directly related to the inertial number introduced later \cite{DaCruz05,GDR04,Jop05,Jop06,Pouliquen06}
	(it is its square actually) as the main quantity relevant to the physics of dry dense granular liquids.
	The inertial number, hereafter denoted $\mathcal{I}$, can be interpreted as the ratio of two characteristic times
	scales associated with grain motions \cite{Cassar05}: At the microscopic scale, the rearrangements can be understood
	as particle motion in a pressure field with time scale $t_m=d\sqrt{n/P}$ ($n$ being the particle's density, $d$
	their diameter, and $P$ the associated pressure),
	whereas at the macroscopic scale they are mainly driven by the imposed shear rate with time $t_M=1/\dot\gamma$.
	The inertial number is then:
	\begin{equation}
		\mathcal{I}=\frac{t_m}{t_M} = \frac{d\,\dot{\gamma}}{\sqrt{P/n}}\ .
	\end{equation}

\begin{figure}
		\begin{center}
			\includegraphics[scale=0.25]{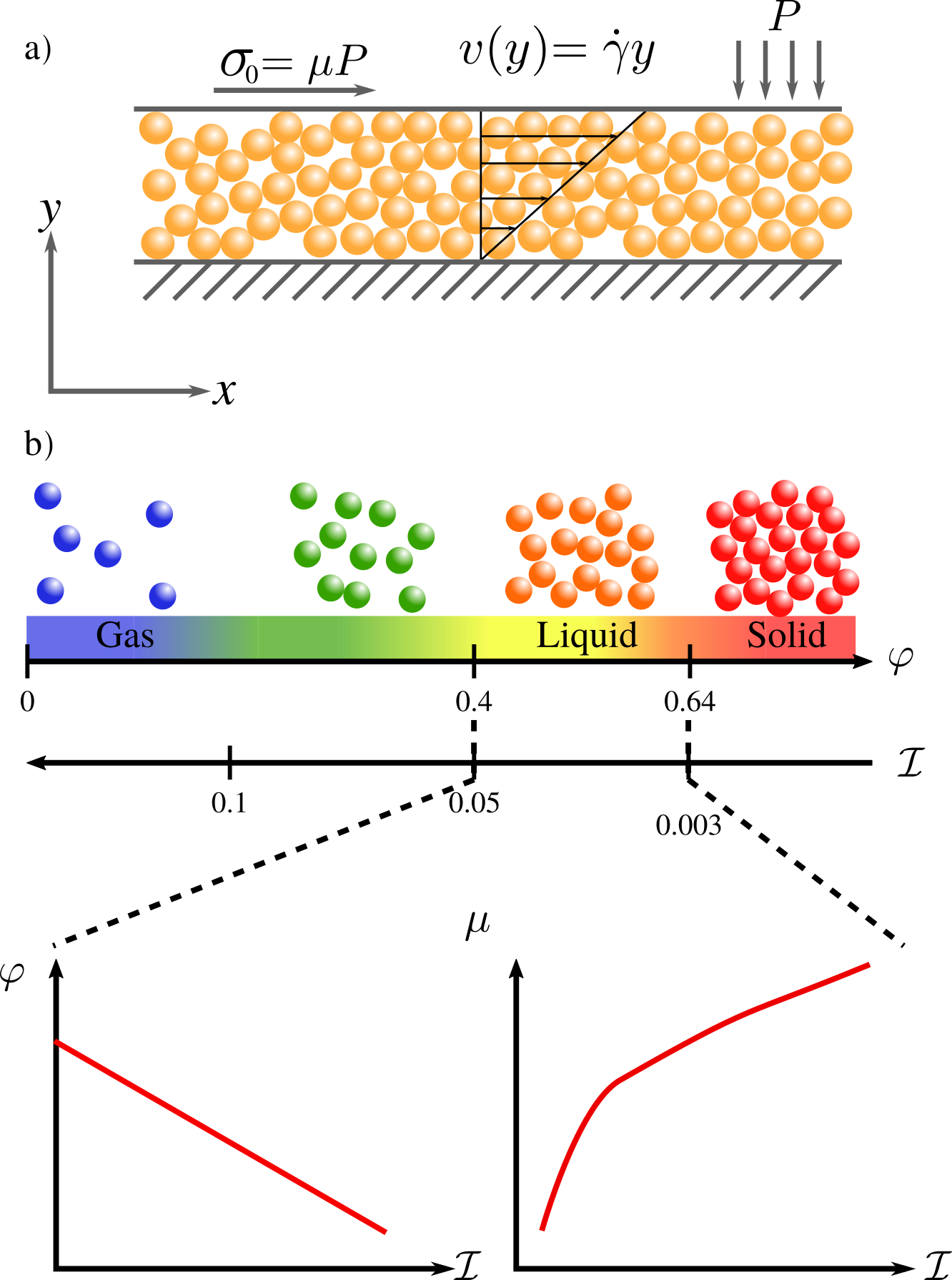}
		\end{center}
		\caption{a) Schematic representation of the plane shear flow geometry.
		b) Schematic representation of the various phases of granular matter as a function of the packing fraction
		$\varphi$ and inertial number $\mathcal{I}$.
		Inside the granular liquid regime, both $\varphi$ and the effective friction $\mu$ depend on $\mathcal{I}$.}
		\label{figReg}
	\end{figure}

	Low values of $\mathcal{I}$ correspond to high pressures and/or small shear rates, they therefore correspond to the solid
	limit, whereas larger $\mathcal{I}$s are more typical of the approach of the gaseous regime
	\cite{DaCruz05,Jop05,Jop06} (see also Fig.~\ref{figReg}).

	A next decisive step has been overcome after the very detailed experimental work \cite{GDR04}:
	In this study, granular liquids were observed in six different flow 
	configurations, a lot of data were collected, and some universal patterns were exhibited.
	The most important result of that study is that the physics of granular liquid flows can be captured by two remarkably simple
	laws depending only on the inertial number $\mathcal{I}$.
	The first law describes the evolution of the packing fraction $\varphi(\mathcal{I})$ and accounts for the dilatancy
	phenomenon; the second one provides a constitutive relation $\mu(\mathcal{I})$, thereby giving to the Mohr-Coulomb parameter
	an explicit dependence on the shear rate.

	This simple framework has subsequently been tested in several experiments and numerical studies \cite{Pouliquen06,Forterre08,Peyneau08,Staron10,Tankeo13,Fullard17,Forterre18,Tapia19},
	and extended to other flow geometries \cite{Lagree11}.
	The mere fact that so many different flow configurations agree with these laws hints that they capture very surely the fundamental
	behavior of granular matter, rather than some experimental artifact.
	Moreover, the framework of this so-called $\mu(\mathcal{I})$ rheology has been extended to the description of the rheology of suspensions of athermal particles
	\cite{Cassar05,Boyer11,Clavaud17,Guazzelli18,Tapia19}.
	Its range of applicability combined with its simplicity is the strongest asset of the $\mu(\mathcal{I})$ rheology.
	Its biggest weakness, however, is its lack of theoretical support \cite{Forterre18}.
	Our study addresses this last point.

	Let us be a little bit more specific about the range of applicability of the $\mu(\mathcal{I})$ rheology.
	Indeed, even within the liquid phase, the physics of granular flows does not seem to be unified.
	More particularly, at high enough densities, a whole range of new phenomena arise, including the jamming physics due to increased importance of friction
	\cite{Ikeda12,Ikeda13,DeGiuli15,DeGiuli16,DeGiuli17a}, discontinuous shear thickening \cite{Wyart14,Brown14}, turbulent-like power laws in the energy spectrum
	\cite{Radjai02,Saitoh16a,Saitoh16,Oyama19}, and the development of creep flows and plastic deformations that make the rheology non-local
	\cite{Pouliquen09,Bocquet09,Frey10,Kamrin12,Hennan12,Bouzid15,Kamrin15,Zhang17,Ozawa18,Nicolas18,Fielding19,Thompson19}.
	Furthermore, it has been estimated in \cite{DeGiuli16} that for $\mathcal{I}\lesssim 10^{-2.5}$, the
	microscopic dynamics of the particles undergoes a transition from a collision to a friction dominated regime.
	Our target regime is therefore densities high enough so that the granular sample can be qualified as liquid,
	but moderate enough so that the $\mu(\mathcal{I})$ rheology can still be applied, so roughly a packing fraction
	in the range $0.40\lesssim\varphi\lesssim0.60$ ,
	and an inertial number $0.05\gtrsim\mathcal{I}\gtrsim 0.003$ (see Fig.~\ref{figReg}).

	In the language of liquid-state theory, the corresponding regime is the one where the competition between the diffusive
	behavior of the particles and the cage effect,
	corresponding to the clogging phenomena due to its neighbors, takes place.
	A natural candidate for a theory in such a regime is therefore mode-coupling theory \cite{Goetze08},
	which is tailored to be efficient precisely in this regime of densities.
	The extension of mode-coupling theory to the rheological context has been made for Brownian suspensions
	\cite{Fuchs02,Henrich07,Brader08,Fuchs09} by the
	use of the integration through transients formula, that we will present below.
	More recently, all this formalism has been adapted to the description of granular fluids \cite{Kranz10,Kranz13,Kranz18,Kranz20}.
	This synthesis will be called Granular Integration Through Transients (GITT) in the following.
	As a particular outcome, it has been shown that such a framework is able to capture all the different qualitative
	flow regimes of granular liquids:
	Newtonian fluid, yielding behavior, and Bagnold regime \cite{Kranz18,Kranz20}.

	In this paper, we show that the $\mu(\mathcal{I})$ rheology is satisfactorily captured by the GITT
	framework, thereby accomplishing a first step in the building of a theoretical description of granular liquids.
	The paper is organized as follows: In a first part, we recall the general properties of GITT 
	and adapt it to the description of the granular rheology
	by adding the description of the evolution of the pressure, that was missing in the previous work \cite{Kranz18,Kranz20}.
	In a second part, we examine GITT predictions for the two laws of the $\mu(\mathcal{I})$ rheology and compare them to
	the experimental ones.
	In particular we show a very reasonable agreement between the GITT values and the modified Mohr-Coulomb law, which
	confers GITT a not only qualitative but also quantitative predictive power for the effective friction.
	Finally we conclude.

\section{Granular Integration Through Transients}

	\subsection{General principles}

	Although the laws of the $\mu(\mathcal{I})$-rheology apply to all kinds of granular flows,
	in the following we will restrict ourselves to the simplest one:
	We consider a set of $N$ inelastic hard spheres of coefficient of restitution $\varepsilon$ submitted to a plane shear flow
	in the plane $(xy)$ (see Fig.~\ref{figReg}).
	If the packing fraction $\varphi$ is not too low, the velocity profile imposed by the shear is
	linear \cite{GDR04}, and can be described as $\mathbf{v}=\kappa\cdot\mathbf{r}$,
	with $\kappa_{ij}=\dot{\gamma}\,\delta_{ix}\delta_{jy}$, where $\dot{\gamma}$ is the shear rate.
	The relation Tr$(\kappa)=0$ ensures incompressibility.
	Although granular systems are out-of-equilibrium systems, a granular temperature $T$ can be defined
	from the second	moment of their velocity distribution, by analogy with thermal systems.

	In order to characterize our liquid state, as discussed in the introduction, we must ensure that the density is sufficiently
	large, so that individual two-body collisions do not dominate, and that the dynamics is fast enough so that the granular
	packing flows on a reasonable time scale.
	This can be characterized by the use of two dimensionless numbers.
	The first one, the Peclet number Pe, compares the efficiency of advection and diffusion in the motion of particles.
	In sheared granular fluids Pe$=\dot{\gamma}/\omega_c$ is the ratio of the shear rate and the collision frequency.
	Pe$ \ll 1$ ensures that collisions are frequent enough at the scale of the applied shear so that the system is in the liquid,
	rather than the gaseous phase.
	When the liquid becomes denser, however, diffusion is strongly hampered by the cage effect; we need another dimensionless
	number to localize the system on the liquid--solid axis.
	This number, called the Weissenberg number Wi, can be written as Wi$=\dot{\gamma}\tau$, where $\tau$ is a typical
	time scale associated to the structural relaxation in the granular liquid \cite{foot2}.
	If $\dot{\gamma}$ is low enough, structural relaxations are not altered by the presence of shear, but at higher $\dot{\gamma}$,
	the shear flow advects the particles thereby breaking the cages formed by their neighbors.
	Whenever Wi$>1\gg$Pe, the physics of the system is governed by a strong competition between structural relaxation
	and shear advection.

	The description of the effect of shear poses a problem in as much as it distorts the phase space available to the particle in a non trivial way.
	A cure to this issue is given by the so-called Integration Through Transients (ITT) formalism \cite{Fuchs02} which 
	relates averages in the sheared system to averages in a reference, quiescent system where no shear is applied
	and therefore statistical averages are easier to compute (see \cite{Fuchs09} for details), at the
	price of introducing an integral over the transient evolution of the system (hence the name).
	For the microscopic stress tensor $\sigma_{\alpha\beta}$ for example, the ITT formula yields:
	\begin{equation}
	\label{eqITT}
		\left<\sigma_{\alpha\beta}\right>^{(\dot\gamma)} = \left<\sigma_{\alpha\beta}\right>_0
		- \int_0^{+\infty}dt\,\left<\frac{\dot{\gamma}\sigma_{xy}}{T}\,\sigma_{\alpha\beta}(t)\right>_0\:,
	\end{equation}
	where we used the convention $k_B=1$, and the particle's mass is $m=1$; $\left<.\right>^{(\dot{\gamma})}$ denotes averages
	in the sheared system,
	$\left<.\right>_0$ denotes averages in the reference state, and $\sigma$ is the stress tensor.
	The term $\dot{\gamma}\sigma_{xy}/T$ is the operator that relates the unsheared reference system to the
	sheared system we want to study.
	Eq.~(\ref{eqITT}) is also called the generalized Green-Kubo relation for the stress tensor.
	This equation represents a stress-strain rate relation, which is precisely what rheology aims to describe.
	Because granular liquids are dissipative, it is important that a source of energy is present in the
	reference system, in order to avoid having a static reference state, where the GITT formalism cannot be applied.
	Therefore, we chose as a reference system a state where no shear stress is present, but an unspecified external source
	of energy maintains the fluid at the same temperature as the real sheared state (the reader is referred to \cite{Kranz20}
	for details).

	In their original works \cite{Fuchs02,Fuchs03,Fuchs09}, Fuchs and Cates introduced this formalism to discuss the
	rheology of colloidal suspensions.
	In that case, transforming averages in the sheared system into averages in the unsheared system amounts to going from out-of-equilibrium averages
	to averages for a system in thermal equilibrium, with Maxwell-Boltzmann velocity distribution.
	For granular liquids, however, further approximations are needed (the reader is referred to \cite{Kranz20} for details).
	Indeed, because of dissipation, the quiescent state is already out-of equilibrium.
	The evolution we describe is therefore connecting two out-of-equilibrium steady states, and for practical reasons similar kinds of hypotheses
	are needed to be able to evaluate the average inside the time integral, so that situations too far from steady states are forbidden.
	Note that in the original use of ITT for colloidal suspensions, situations too far away from equilibrium were not considered either \cite{Brader08}.
	
	Another necessary specification in the case of granular systems is that since both the sheared and the unsheared dynamics are dissipative, the stress
	operator in Eq.~(\ref{eqITT}) is defined with an elastic collision operator \cite{Kranz20}, and will therefore be denoted $\sigma^{el}$ in the following.

	Next, an approximation of the stress correlation in Eq.~(\ref{eqITT}) is needed.
	Since the main effect driving its evolution is the slow dynamics generated by the cage effect (at least in the liquid phase where the density is high enough),
	a natural candidate is Mode-Coupling Theory (MCT).
	It consists of two main steps: first everything is projected onto pairs of density fluctuation modes, then four-point density fluctuations
	are factorized into a product of two two-point fluctuations.
	The result is \cite{Fuchs02,Kranz20} :
	\begin{equation}
	\label{eqITTVW}
		\begin{split}
			\left<\sigma_{\alpha\beta}\right>^{(\dot\gamma)} &= \left<\sigma_{\alpha\beta}\right>_0 \\
			&-\frac{\dot\gamma}{2T} \int_0^{+\infty}dt\int\frac{d^3k}{(2\pi)^3}\,
			\mathcal{V}_{k(-t)}^\sigma \mathcal{W}_{k,\alpha\beta}^\sigma\,\Phi^2_{k(-t)}(t)\:,
		\end{split}
	\end{equation}
	where the integrand now consists of the dynamical structure factor (or transient correlator) $\Phi_k(t)=N\left<\rho_k(t)\rho_{-k}\right>/S_k$,
	and two vertices defined as \cite{Kranz20,foot3}:
	\begin{equation}
		\begin{split}
			& \mathcal{V}_k^\sigma               = N \left<\sigma^{el}_{xy} \big| \rho_k\rho_{-k} \right> \\
			& \mathcal{W}_{k,\alpha\beta}^\sigma = N \left<\rho_k\rho_{-k}  \big| \sigma_{\alpha\beta} \right>/S_k^2 \ .
		\end{split}
	\end{equation}
	Because the system flows, the memory of its initial density distribution is lost at large
	enough times, the transient correlator must therefore relax to 0 in that limit.
	Due to the cage effect, the transient correlator $\Phi_k$ will typically develop a plateau in its relaxation towards 0.
	This plateau is generally smaller at large values of $k$.
	Because in Eq.~(\ref{eqITTVW}) the transient correlator is evaluated in its advected wave vector $\mathbf{k}(-t)$, whose norm is bigger
	than that of the original one, $k(-t) = k\sqrt{1+(\dot{\gamma}t)^2/3}$, the transient correlator gets smaller as $t$ gets larger,
	thereby reproducing the competition between the slow structural relaxation and shear which, through advection, tends to break the cages.
	At some point $\Phi_k(t)$ exponentially relaxes to zero (the system is in a liquid state), so that the integral term never diverges.
	The time evolution of $\Phi_k(t)$ is given by a Mode Coupling equation of motion (see Appendix. \ref{AMCT}, and \cite{Kranz20}
	for more details).

	\subsection{Application to granular rheology}

		By the combination of ITT and MCT, the GITT formalism is built to be efficient in the description of the competition of shear advection and structural relaxation,
		and thus constitutes a natural candidate for the description of the physics of moderately dense granular liquids.
		Its greatest success has been its successful description of the various flow regimes in granular liquids \cite{Kranz20}, as well as the appearance of
		a yield stress at low shear rate and high density.
		In this paper, the study is focused on the description of the physics of the granular steady flow in the case where the only external energy input comes from the
		shear (the granular packing is not fluidized).
		Under such conditions, the flow is always in the Bagnold regime $\sigma\propto\dot{\gamma}^2$ \cite{Bagnold54}.

		The interpretation of the stress tensor as a source term in the momentum conservation equation allows to write the vertices $\mathcal{V}_k^\sigma$
		and $\mathcal{W}_k^\sigma$ as standard correlations of pairs of density and particle's current \cite{Kranz20}, which have been evaluated in
		the case of inelastic hard spheres \cite{Kranz13}.

		Even if $\sigma_{\alpha\beta}$ is a tensor, symmetries as well as the isotropic approximation used here reduce its components to only two
		independent scalars.
		An appropriate way to decompose it, is to use the projectors longitudinal to and transverse to the current wave vector $\mathbf{q}$,
		defined as:
		\begin{subequations}
		\label{eqPLPT}
		\begin{equation}
				P_L^{\alpha\beta}(q) = \frac{q^\alpha q^\beta}{q^2} \,, \\
		\end{equation}
		\begin{equation}
				P_T^{\alpha\beta}(q) = \delta^{\alpha\beta} - P_L^{\alpha\beta}(q)\ .
		\end{equation}
		\end{subequations}
		Hence, the microscopic stress tensor is decomposed as:
		\begin{equation}
		\label{eqSigPLPT}
			\sigma_{\alpha\beta} = \sigma_L\, P_L^{\alpha\beta}(q) + \sigma_T\, P_T^{\alpha\beta}(q)\ ,
		\end{equation}
		with \cite{Kranz20,foot1}:
		\begin{subequations}
		\label{eqSigmaScal}
		\begin{equation}
				\sigma_L = \left(\frac{1+\varepsilon}{2}\right)T\left[-k\,S_k'+S_k-S_k^2\right]\\
		\end{equation}
		\begin{equation}
				\sigma_T = \left(\frac{1+\varepsilon}{2}\right)T\left[S_k-S_k^2\right]\ .
		\end{equation}
		\end{subequations}
		These expressions depend explicitly on the restitution coefficient $\varepsilon$.
		Note the in the case of vertex $\mathcal{V}_k^\sigma$, the operator in the average is the elastic stress tensor.
		The formulas Eq.~(\ref{eqSigmaScal}) can be applied, but with $\varepsilon = 1$.

		The macroscopic shear stress $\sigma_0$ is derived from the average of
		the off-diagonal component $\sigma_{xy}$.
		The associated ITT vertex is thus:
		\begin{equation}
		\label{eqWs}
			\mathcal{W}_{k,xy}^\sigma = \hat{k}_x \hat{k}_y (\sigma_L - \sigma_T)
			= -\hat{k}_x \hat{k}_y \left(\frac{1+\varepsilon}{2}\right) T\,k\,S_k' \ ,
		\end{equation}
		where $\hat{.}$ is used to represent a normalized vector.

		For the pressure however, the diagonal components are involved.
		Defining the microscopic pressure as Tr$(\sigma_{\alpha\beta})/3$,
		the vertex $\mathcal{W}_k^P = (\mathcal{W}_{k,xx}^{\sigma} + \mathcal{W}_{k,yy}^{\sigma} +\mathcal{W}_{k,zz}^{\sigma})
		/3$ becomes:
		\begin{equation}
		\label{eqWp}
			\begin{split}
				\mathcal{W}_k^P &= \frac{\hat{k}_x\hat{k}_x + \hat{k}_y\hat{k}_y +\hat{k}_z\hat{k}_z}{3} (\sigma_L - \sigma_T) + \sigma_T \\
				&= -\frac{\hat{k}_x\hat{k}_x + \hat{k}_y\hat{k}_y +\hat{k}_z\hat{k}_z}{3} \left(\frac{1+\varepsilon}{2}\right) T\,k\,S_k' \\
				\ &+\left(\frac{1+\varepsilon}{2}\right) T \big[S_k - S_k^2\big]\ .
			\end{split}
		\end{equation}
		Since the diagonal components do not depend solely on the stress component differences, the vertex now involves a second term.
		
		The next step consists in injecting the expressions of the vertices in the ITT equation Eq.~(\ref{eqITTVW}), and performing the angular integrals.
		At this stage, it is interesting to introduce the following auxiliary functions that carry the remaining wave vector dependence,
		\begin{subequations}
		\label{eqF}
		\begin{equation}
			F_{1}(k,t) = -k^4\,\dot{\gamma}T\left(\frac{1+\varepsilon}{2}\right)\Phi^2_{k(-t)}\frac{S'_{k(-t)}S_k'}{S_k^2}\\
		\end{equation}
		\begin{equation}
			F_{2}(k,t) = -k^3\,\dot{\gamma}T\left(\frac{1+\varepsilon}{2}\right)\Phi^2_{k(-t)}\frac{S'_{k(-t)}}{S_k^2}(S_k^2 - S_k)\ ,
		\end{equation}
		\end{subequations}
		so that finally,
		\begin{subequations}
		\label{eqRheo}
		\begin{equation}
		\label{eqRheoA}
				\sigma_0 = \frac{1}{60\pi^2}\int_0^{+\infty}dt\frac{1}{\sqrt{1+\frac{(\dot{\gamma}t)^2}{3}}}
				\int_0^{+\infty}dk\,F_{1}(k,t)
		\end{equation}
		\begin{equation}
		\label{eqRheoB}
			\begin{split}
				& P(\dot\gamma) = P(\dot{\gamma} = 0) \\
				&\quad + \frac{1}{36\pi^2}\int_0^{+\infty}dt\frac{(\dot{\gamma}t)}{\sqrt{1+\frac{(\dot{\gamma}t)^2}{3}}}
				\int_0^{+\infty}dk\,F_{1}(k,t)\\
				& \quad + \frac{1}{12\pi^2}\int_0^{+\infty}dt\frac{(\dot{\gamma}t)}{\sqrt{1+\frac{(\dot{\gamma}t)^2}{3}}}
				\int_0^{+\infty}dk\,F_{2}(k,t)\ ,
			\end{split}
		\end{equation}
		\end{subequations}
		where $P$ is the macroscopic pressure, and
		the shear-rate dependence of the pressure has been made explicit (the shear-stress is 0 in the unsheared system).

		The equation (\ref{eqRheoA}) gives access to the shear stress $\sigma_0$, and thus also to the viscosity of the granular
		fluid $\eta = \sigma_0/\dot\gamma$, as well as its Bagnold coefficient $B = \sigma_0/\dot\gamma^2$.
		Their properties have been investigated in \cite{Kranz20}.

		The pressure equation, Eq.~(\ref{eqRheoB}) has a quite similar structure, with some subtleties: The different combinations of $\mathbf{k}$-coordinates
		involved in the pressure vertex, Eq.~(\ref{eqWp}), bring in a $t$ in the numerator of the first factor of the integrand, thereby reducing its ability
		to dampen the large time contributions, so that the pressure is much more sensitive to late time events than the shear stress;
		then as we discussed above the pressure vertex is composed of two terms instead of just one, the second of which has a different $\mathbf{k}$-dependence;
		and last but not least, the ITT integral only gives access to the pressure \textit{correction} brought by the slow-down of the dynamics.
		In order to be able to express the full pressure evolution in the granular fluid, it is necessary to know the
		pressure in the unsheared state.
		This is what we examine in the next section.

	\subsection{An equation of state for inelastic hard spheres}

		The first challenge in this problem is to handle the effects of dissipation.
		However, it turns out to be quite easy to overcome.
		Indeed, in a thorough study involving both simulation data and computations from Enskog theory of granular gases
		\cite{Lutsko04}, Lutsko	showed that the pressure displays no dependence on the coefficient of restitution,
		a result that does not seem to deteriorate when the density is increased.
		Although this may be deemed surprising at first, let us recall that in the framework of rheology, we are mostly interested in out-of-equilibrium steady
		states, and therefore in particular the granular fluid does not cool down.
		The dissipation is compensated for at the macroscopic scale by the shearing mechanism.
		It is to be noted that we always consider a setup at fixed shear rate $\dot\gamma$, without asking how much energy is needed to maintain the steady flow.

		The problem of determining the pressure of the unsheared fluid thus reduces to the computation of the equation of state of a hard-sphere fluid,
		a widely studied subject, that still received a fair amount of attention in the last few years (see \cite{Kofala04,Clisby06,Bannerman10,Solana15,
		HansenGoos16,Nikolaev17,Pieprzyk19,Tian19} and references therein).
		Indeed, if the Carnahan-Starling expression is often considered as the best compromise between simplicity and precision, it is not the most
		accurate in our regime of interest ($\varphi\gtrsim 0.4$).
		In a recent review \cite{Tian19}, Tian \textit{et al.} made a comparison of all the state of the art
		expressions of the equation of state with numerical simulation results,
		using three different types of precision criteria.
		In this work, one ansatz appeared to perform much better than all the other: the modified Kofala expression derived by S. Pieprzyk \textit{et al.} \cite{Pieprzyk19}.

\begin{figure}
			\begin{center}
				\includegraphics[scale=0.5]{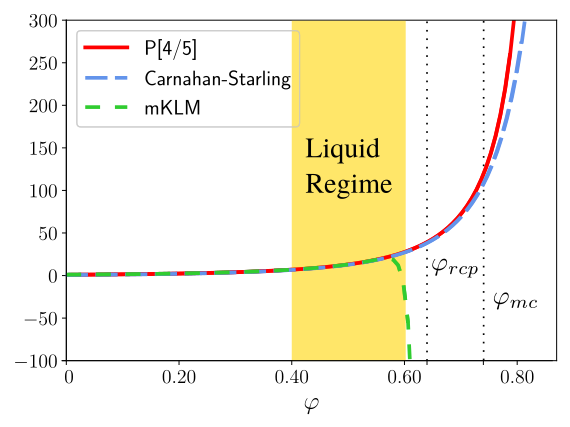}
			\end{center}
			\caption{Comparison of three analytical equations of state.
			mKLM refers to the modified Kofala ansatz of S. Pieprzyk \textit{et al.} \cite{Pieprzyk19},
			which can be shown to fail inside the granular liquid regime, and is therefore not suitable
			for our purpose.
			P[4/5] is the Pad\'e expression of Clisby and Mc Coy \cite{Clisby06} that we have used.
			The fact that these ansatz are continuous at random close packing $\varphi_{rcp}\simeq 0.64$ and
			at maximal compacity $\varphi_{mc}\simeq 0.74$ is an indication that none of the present
			formulas are adapted to the high density regime.}
			\label{figEos}
		\end{figure}

		However, all these equations of state where adjusted on the sector $\varphi\lesssim 0.5$.
		As a result, some of them, including that of \cite{Pieprzyk19} present severe problems at higher packing fraction, such as negative pressure,
		or even divergences towards $-\infty$ at finite packing fractions (see Fig.~\ref{figEos}).
		Hence, such solutions are not suitable to our purpose.

		We therefore required, among all solutions presented in \cite{Tian19},
		the optimal solution regarding the following criteria:
		(i) the solution must be the best possible on the largest range of packing fractions studied in \cite{Tian19}, and (ii) the pressure should never become negative.
		This solution is the Pad\'e P[4/5] of Clisby and Mc Coy \cite{Clisby06}.
		It also presents the nice property of diverging to $+\infty$ at a finite packing fraction $\varphi\simeq0.85$ although a bit too far inside the solid regime
		(such a high packing fraction cannot be reached by hard spheres).
		Although this ansatz is expected to perform rather well in our problem, it pinpoints a crucial feature of pressure computations (and any other related 
		quantity) in theoretical frameworks: The evaluation of the unsheared pressure is an irreducible source of precision loss that cannot
		be easily overcome, even in the case of hard spheres.
		Even if we were able to compute the ITT correction quasi-exactly, the presence of $P(\dot{\gamma}=0)$ in Eq.~(\ref{eqRheoB}) impedes a high precision
		determination of the full pressure, which turns out to be the relevant quantity when it comes to rheology.

		Finally, in order to ensure the overall consistency of the framework, we choose to adapt the structure factor --- whose role in Eq.~(\ref{eqRheo}) is of
		paramount importance --- to the chosen equation of state.
		This is possible within the so-called Rational Fraction Approximation (RFA) method: With such method the structure factor is approximated by a Pad\'e
		approximant, whose coefficients are given by physical constraints imposed on the system.
		In the hard-sphere case, the minimal set of assumptions gives the Percus-Yevick structure factor, but Robles
		\textit{et al.} showed in \cite{Robles97}
		how from a given equation of state it is possible to build with RFA a structure factor that (i) is consistent with this equation of state,
		and (ii) is thermodynamically consistent (indeed, in the case of the Percus-Yevick solution the pressure and compressibility route do not give the
		same equation of state \cite{Hansen06}, and it was shown in \cite{Coquand19} how both equations of state are related respectively to the large
		wave vector and small wave vector sectors of the structure factor).

	\subsection{The pressure of granular liquids}
	
\begin{widetext}

		\begin{figure}
			\begin{center}
				\includegraphics[scale=0.7]{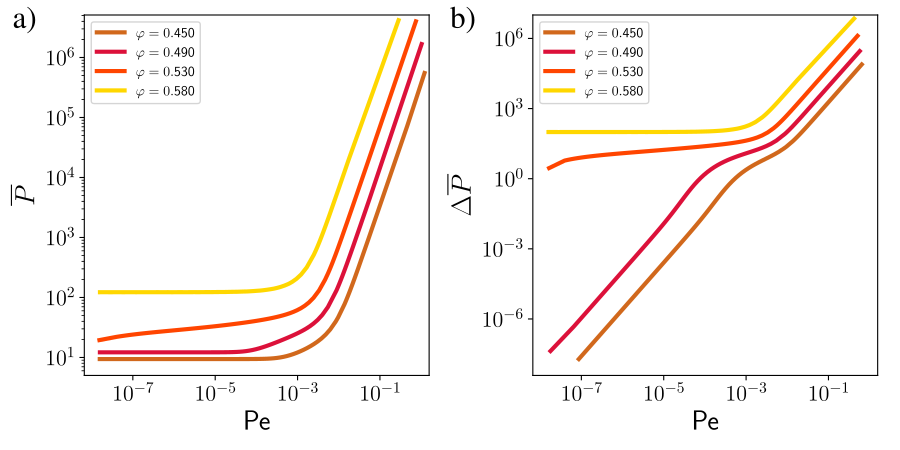}
			\end{center}
			\caption{Evolution of the dimensionless pressure $\overline{P} = P/nT$ of the granular liquid
			as a function of the Peclet number for various packing fractions \cite{foot5}.
			a) total dimensionless pressure. b) contribution from the ITT integral
			$\Delta \overline{P} = \overline{P} - \overline{P}(\dot{\gamma}=0)$.
			All values correspond to a coefficient of restitution $\varepsilon=0.85$.}
		\label{figPr}
		\end{figure}

		\end{widetext}

		We can now discuss the full pressure $P$ given by Eq.~(\ref{eqRheoB}).
		The results are displayed in Fig.~\ref{figPr}.
		
		In order to understand better the evolution of $P$, it is interesting to compare it to the evolution of the ITT correction
		$\Delta P = P - P(\dot{\gamma}=0)$, which can be understood as a competition between structural relaxations and shear advection.
		
		Let us first analyse the lowest densities --- $0.40\lesssim\varphi\lesssim \varphi_g$, where $\varphi_g\simeq0.52$ \cite{Kranz18}
		is the location of the MCT granular glass transition in the unsheared system.
		In this low density region the cage-effect is weak,  and the structural time $\tau$ is given by diffusion in the liquid.
		In that case, Wi$\simeq$Pe.
		At the lowest Pe's, Wi is thus also small and structural relaxations dominate.
		Since the granular liquid is here in the Newtonian regime \citep{Kranz18}, the contribution brought by structural relaxations
		is small, and $\Delta P << P$ (see Fig.~\ref{figPr}).
		Then, as Pe increases, so does Wi, and for high enough $\dot{\gamma}$ shear advection dominates; this corresponds to the
		Bagnold regime.
		In the Bagnold regime, $\Delta P \simeq P$.

		For higher packing fractions ($\varphi\gtrsim\varphi_g$), the behavior in the Bagnold regime is not different.
		At lower Pe's, however, the system enters the yielding regime \cite{Kranz18}: as Pe is decreased, the weaker shear is less
		efficient to break the cages, and the structural relaxation time $\tau$ is significant.
		Therefore, Wi$\gg$Pe, even at low $\dot{\gamma}$.
		As a result, $\Delta P$ is still comparable to $P$, even in the limit Pe$\rightarrow0$.
		
		All in all, the pressure in granular fluids is significantly altered by shear advection, except in the Newtonian regime
		$\varphi<\varphi_g$, Pe$<10^{-3}$.

\section{The $\mu(\mathcal{I})$ rheology}

	\subsection{Presentation}

		The rheological behavior of the granular liquid is captured by two simple laws which have been obtained by fitting a huge data set from various
		experiments \cite{GDR04}~:
		first, the packing fraction behaves as \cite{DaCruz05,Pouliquen06}:
		\begin{equation}
		\label{eqPhiexp}
			\varphi(\mathcal{I}) = \varphi_c + (\varphi_m-\varphi_c)\,\mathcal{I}\ ,
		\end{equation}
		where typically $\varphi_m\simeq0.4$ \cite{Forterre08}.
		The upper bound of the packing fraction $\varphi_c$ depends on friction: For frictionless particles, it is expected to be
		equal to the random closed packing value $\varphi_c=0.64$, whereas for frictional particles, typical values are $\varphi_c\simeq0.58$ \cite{Tapia19}.
		The law Eq.~(\ref{eqPhiexp})  accounts for the dilatancy phenomenon observed in granular flows.
		Depending on whether the volume of the system or its pressure is fixed, it can be replaced by an equation giving $P(\mathcal{I})$.
		Here, we chose to discuss the imposed pressure setup implemented in most experiments.
		
		The second law is a modified Mohr-Coulomb criterion:
		\begin{equation}
		\label{eqMuexp}
			\mu(\mathcal{I}) = \mu_1 + \frac{\mu_2 - \mu_1}{\mathcal{I}_0/\mathcal{I}+1} \ .
		\end{equation}
		Typical values for monodisperse glass beads are $\mu_1\simeq0.38\simeq\tan(21^\circ)$, $\mu_2\simeq0.65\simeq\tan(33^\circ)$ and $\mathcal{I}_0\simeq0.3$ \cite{Forterre08}.
		Such a law thus describes a decrease of the effective friction for denser flows, even though it remains a mild effect, as can be assessed from the
		values of the corresponding angles.

		Obviously, the really dilute limit in which a proper granular gas is considered does not fit well our description.
		For example, the previously exhibited
		mean velocity profile generated by the shear does not hold anymore \cite{GDR04}.
		The laws of the so-called $\mu(\mathcal{I})$ rheology are designed to describe the physics dominated by the competition
		between solid-like and liquid-like behaviors, which for granular matter corresponds to the liquid phase.
		In the $\mathcal{I}\simeq1$ limit, kinetic theory should be preferred.
		When $\mathcal{I}$ becomes low, typically $\mathcal{I}\lesssim 10^{-3}$, friction dominates \cite{Staron10,DeGiuli15,DeGiuli16} and the simple
		rheology fails.
		These limits correspond to the range of packing fractions introduced above.

		The great success of the $\mu(\mathcal{I})$ laws Eq.~(\ref{eqPhiexp}) and Eq.~(\ref{eqMuexp}) comes from their ability to describe the phenomenology
		of granular liquid flows in many different setups \cite{GDR04} such as simple shear, Couette cell, flow down a heap, rotating drum experiment, 
		but also the collapse of a granular column \cite{Lagree11}, and a number of geophysical phenomena (see \cite{Pahtz20} and references therein).
		Hence, the chances are high that, although they are simply fitted on some experiments without theoretical support so far, these laws are able
		to capture at least part of the fundamental behavior of granular liquids.

		In the following sections, we discuss how the laws Eq.~(\ref{eqPhiexp}) and Eq.~(\ref{eqMuexp}) agree with the GITT formalism.

	\subsection{GITT in the Bagnold regime}

		The granular liquid being a dissipative system, the existence of an out-of-equilibrium steady state
		requires a balance between the power injected into the system, and the dissipated power.
		In the Bagnold regime, there is no source of power other than the shear heating.
		The power balance equation thus becomes \cite{Kranz20}:

		\begin{equation}
		\label{eqBag}
			\sigma\dot\gamma = n \Gamma_d\,\omega_c\,T\ ,
		\end{equation}
		where $\Gamma_d(\varphi,\varepsilon)$ is the dimensionless dissipation rate.

		The evolution of $\varphi$ and $\mu$ with $\mathcal{I}$ can then be deduced from the GITT equations
		Eq.~(\ref{eqRheo}).
		Indeed, with a given packing fraction, restitution coefficient, and Peclet number --- playing here the role
		of dimensionless shear rate --- these equations determine $\sigma_0$ and $P$.
		Moreover, in the Bagnold regime, Pe is fixed by Eq.~(\ref{eqBag}) once $\varphi$ and $\varepsilon$
		are set \cite{Kranz20,foot4}.
		By computing $\sigma_0$ and $P$ thanks to the GITT equations (\ref{eqRheo}) for various $\varphi$s at a given
		$\varepsilon$, it is possible to get numerical estimates of $\varphi(\mathcal{I})$ and $\mu(\mathcal{I})$.
		Those are displayed on Figs.~\ref{figphinofit} and \ref{figmu}.

		Note that in Eq.~(\ref{eqRheoB}), to a given set $(\varphi,\varepsilon,$Pe$)$ corresponds a unique value of $P$.
		Therefore equivalent results would have been obtained by considering the set $(P,\varepsilon,$Pe$)$ and
		varying $P$.

	\subsection{The packing fraction law -- Dilatancy effect}

\begin{figure}
			\begin{center}
				\includegraphics[scale=0.5]{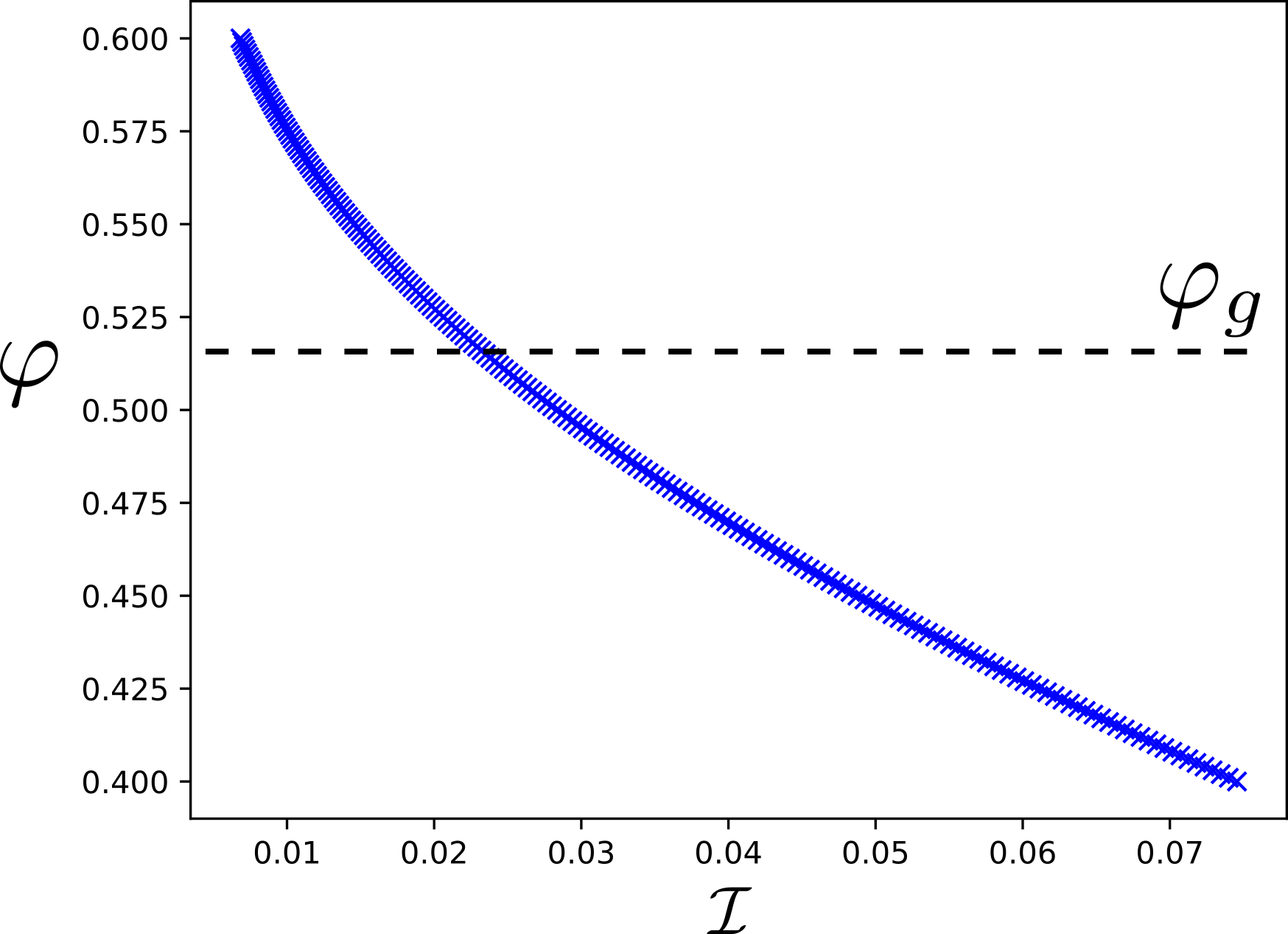}
			\end{center}
			\caption{Evolution of the packing fraction $\varphi$ as a function of the inertial number $\mathcal{I}$
			within GITT for $\varepsilon = 0.85$.
			Above the granular glass transition packing fraction $\varphi_g\simeq0.52$, the behavior is clearly not linear anymore.}
			\label{figphinofit}
		\end{figure}

		The GITT predictions for the evolution of the packing fraction with the inertial number is shown on Fig.~\ref{figphinofit}.
		It is striking that the behavior is not linear throughout the whole range of $\mathcal{I}$, 
		as expected from Eq.~(\ref{eqPhiexp}), even though the regime $\mathcal{I}\lesssim10^{-3}$ is not reached.

		Many reasons can come to mind to explain this inconsistency between theory and experiment.
		First, it is a known problem of MCT, that we used in GITT, that it tends to overestimate the cage effect in the fluid if the fluid becomes too dense.
		Therefore, some caution has to be taken when looking at predictions of packing fractions close to 60\%.
		Moreover, around $\mathcal{I}=10^{-3}$, it is expected that the flow enters a friction dominated regime \cite{Staron10,DeGiuli16,DeGiuli17a,DeGiuli17b},
		which effects should begin to come into play in the lowest $\mathcal{I}$s of our data, but are completely absent of our frictionless model.

		However, the extent to which such failures are responsible for the behavior observed on Fig.~\ref{figphinofit} is not so clear.
		In particular, the transition to a non-linear regime, imposed to reach the jamming transition \cite{DeGiuli16} is not
		observed experimentally, to the best of our knowledge, even in very recent studies \cite{Tapia19} (the furthest data points have $\mathcal{I}\simeq5\times 10^{-2}$),
		which rules out the influence of friction as a convincing explanation of the discrepancies between experiments and GITT.
		There is indeed another source of error, which is the estimation of the pressure in the quiescent hard sphere fluid.

		Let us elaborate below a simple mechanism through which this issue alone could account for the observed phenomenon.
		A striking feature of the evolution of $\varphi$ in Fig.~\ref{figphinofit} is its superlinear character above the granular glass transition
		$\varphi_g$ \cite{Kranz10,Kranz13}.
		This is indeed quite surprising since at low $\mathcal{I}$ the corrections to the linear behavior imposed by the approach of jamming are typically
		sublinear \cite{DeGiuli16,DeGiuli17a,DeGiuli17b}.
		As explained above, it is expected that our hard spheres equation of state underestimates the true value of the pressure, which is all the more
		problematic that its contribution is typically half of the total pressure at the highest packing fractions we investigated.
		Note also that the pressure estimate is all the poorer that $\mathcal{I}$ is low, so that the resulting distortion of the $\mathcal{I}$
		is a non-linear transformation.

\begin{figure}
			\begin{center}
				\includegraphics[scale=0.5]{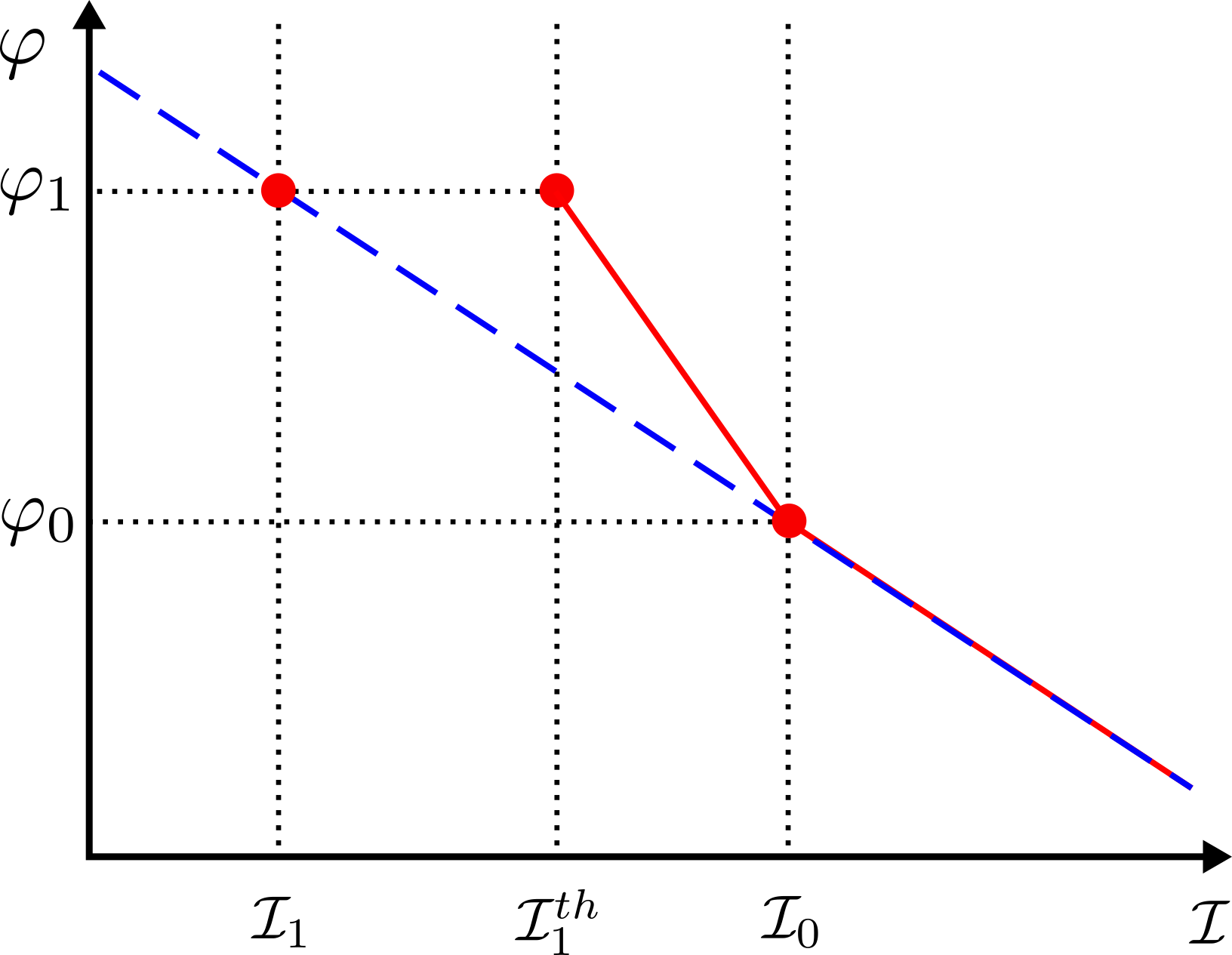}
			\end{center}
			\caption{Schematic evolution of $\varphi$ with $\mathcal{I}$.
			The dashed line is the experimental curve, the solid line the theoretical one.
			The deformation of the $\mathcal{I}$ axis in the theory gives the illusion of a superlinear evolution.}
			\label{figphierr}
		\end{figure}

		Suppose that the linear behavior of the packing fraction with $\mathcal{I}$ is well reproduced up to $\mathcal{I}_0$, where it is $\varphi_0$ (see Fig.~\ref{figphierr}).
		If we further choose $10^{-3}<\mathcal{I}_1<\mathcal{I}_0$, the evolution of the packing fraction of the real system is expected to be linear on
		$[\mathcal{I}_1;\mathcal{I}_0]$.
		We define $\varphi_1=\varphi(\mathcal{I}_1)$ (see Fig.~\ref{figphierr}).
		Now, let us suppose that our theoretical model reproduces exactly the real value $\varphi^{th}_1=\varphi_1$ for the given shear rate and granular temperature,
		and that the only source of error is the determination of the equation of state of elastic hard spheres (this is obviously overstated, but let us assume
		it is true for the sake of the argument).
		Thus, because the theoretical pressure on this point is underestimated compared to the real value, the corresponding inertial number, which varies as the
		inverse of the pressure's square root, is such that $\mathcal{I}_1^{th}>\mathcal{I}_1$.
		As a result, the evolution of the theoretical estimate $\varphi^{th}$ becomes superlinear, as can be seen on Fig.~\ref{figphierr}.
		Iterating this procedure, and taking into account the fact that the pressure estimate worsens when lowering $\mathcal{I}$, the resulting shape
		becomes similar to that of Fig.~\ref{figphinofit}.

		Of course, it would be better to have a more precise numerical estimate of the pressure underestimation to be able to conclude about the extent to which
		the exposed mechanism alone is responsible for the non-linear evolution of $\varphi$ with $\mathcal{I}$.
		However, as explained above, we already used the best available pressure evolution in this packing fraction range.

		All in all, GITT only reproduces a linear evolution of the packing fraction with the inertial number up to packing fractions of the order
		of those where the granular glass transition occurs.
		This is probably the consequence of our lack of a precise equation of state to determine the pressure of the quiescent fluid.

		\subsection{The effective friction law -- Mohr-Coulomb criterion}

\begin{figure}
				\begin{center}
					\includegraphics[scale=0.5]{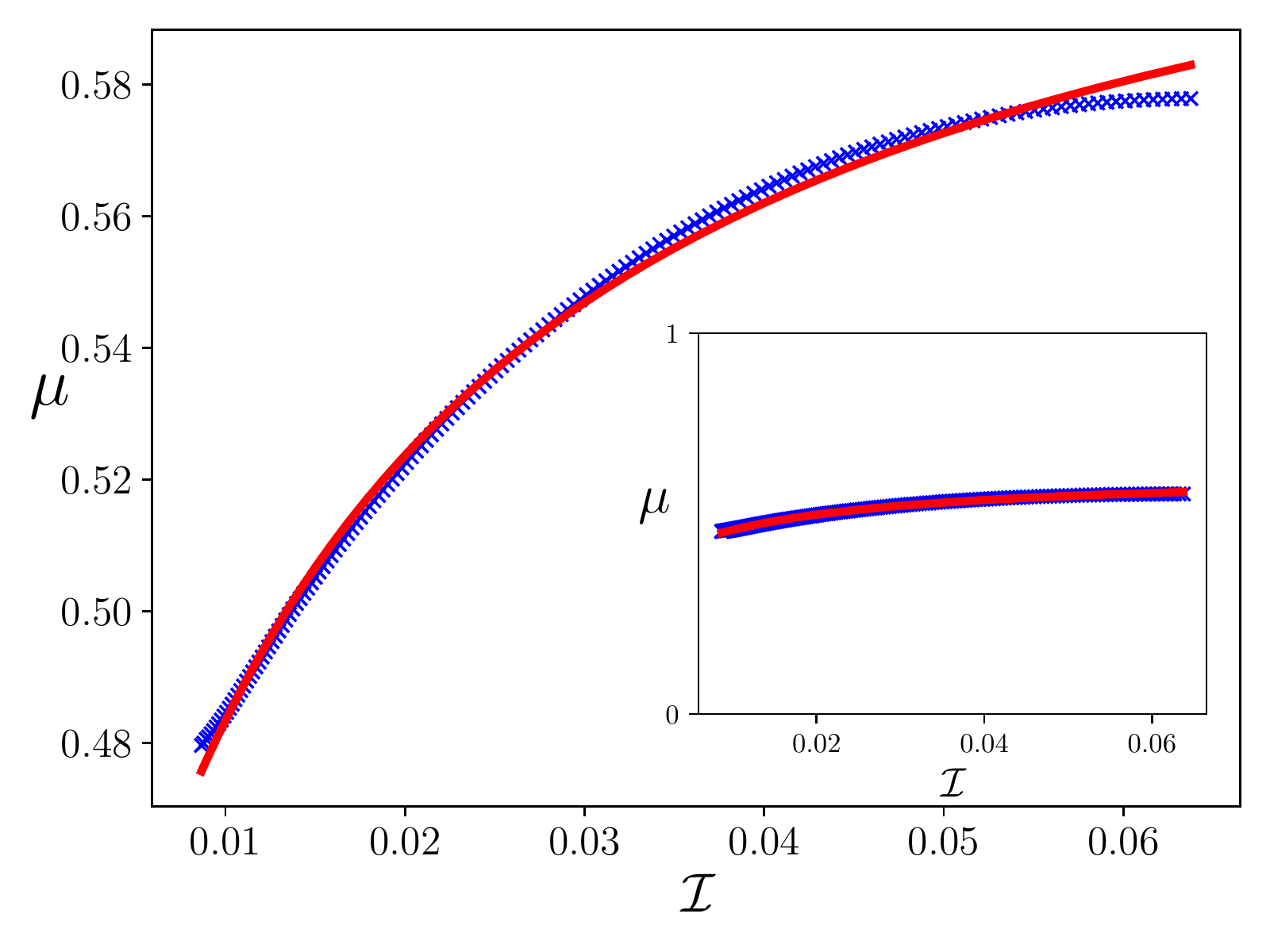}
				\end{center}
				\caption{Evolution of the effective friction $\mu$ as a function of the inertial number $\mathcal{I}$
				within GITT (crosses) for a coefficient of restitution $\varepsilon = 0.85$.
				The solid line is a fit with respect to the form Eq.~(\ref{eqMuexp}).
				The inset shows that the variations of $\mu$ with $\mathcal{I}$ remain very mild.}
				\label{figmu}
			\end{figure}

			The results for the effective friction coefficient are presented in Fig.~\ref{figmu}.
			The solid line represents the best fit with respect to the experimental law Eq.~(\ref{eqMuexp}).

			Contrary to the packing fraction, the evolution of $\mu(\mathcal{I})$ fits quite convincingly with the experimental law.
			The fitting parameters are presented in Table.~\ref{tabmu}, along with some other values used or measured in previous works.

			Let us first examine the parameter $\mathcal{I}_0$ that gives the typical scale separating the two regimes with roughly constant
			effective friction ($\mu\simeq\mu_1$ and $\mu\simeq\mu_2$).
			The estimate from our work is one order of magnitude weaker than the typical value measured in \cite{Forterre08}.
			It is to be noted that $\mathcal{I}_0$ is supposed to depend on the type of material, what is clearly seen when comparing with the \cite{Fullard17}
			value (which is much closer to ours).
			However, from our above discussion about the precise value of $\mathcal{I}$ in our model, it could also be that $\mathcal{I}_0$ is used as a
			renormalized scale that absorbs most of the discrepancies between our $\mathcal{I}$ and the experimental one, thereby enabling better
			quality predictions.

			\begin{table}
				\begin{tabular}{ccccccc}
					\hline
					\hline
					                  & & $\mu_1$ & & $\mu_2$ & & $\mathcal{I}_0$ \\
				 	\hline
				 	\hline
					This work         & & $0.39$  & & $0.62$  & & $0.015$         \\
					\cite{Forterre08} & & $0.38$  & & $0.65$  & & $0.3$           \\
					\cite{Lagree11}   & & $0.32$  & & $0.60$  & & $0.4$           \\
					\cite{Fullard17}  & & $0.57$  & & $0.82$  & & $0.05$          \\
				 	\hline
				\end{tabular}
				\caption{Comparison of the fitting parameters of the friction law, Eq.~(\ref{eqMuexp}),
				from various works.}
				\label{tabmu}
			\end{table}

			Then, the two limiting values of the effective friction $\mu_1$ and $\mu_2$ are remarkably close to their experimental values for
			glass beads \cite{Forterre08}.
			This is all the more interesting that our model is frictionless.
			Consequently, our result support the fact that although $\mu_1$ corresponds to the asymptotic value of $\mu$ in the solid limit,
			it is not related at all to the interparticle friction.
			Such a result has already been observed in simulations of frictionless spheres \cite{Peyneau08}, and then supported by further studies
			(see \cite{Pahtz20} and references therein).
			Indeed, as exposed in \cite{Clavaud17}, steric effects alone are enough to explain the ability of hard sphere packings to sustain a certain
			tilt before beginning to flow; and there is no doubt that steric effects are well captured by GITT.

			Beyond that, the satisfactory agreement between the GITT model and the experimental values of the effective friction coefficients
			supports the idea that the Mohr-Coulomb behavior of granular liquids should be understood as a collective effect at the macroscopic
			scale rather than something linked to a specific mechanism at the grain level.
			This would explain not only why the interparticle friction, but other characteristics such as polydispersity seem to play a very
			minor role in the $\mu(\mathcal{I})$ regime\cite{Voivret09,Pahtz20}.
			Additionally, it is probably one of the reasons why similar rheological behaviors are observed in dense suspensions of
			non-Brownian particles \cite{Boyer11,Forterre18,Guazzelli18,Tapia19}.

			Let us stress again, however, that our results are not expected to hold up to arbitrarily
			low $\mathcal{I}$s.
			For $\mathcal{I} \lesssim 10^{-2.5}$, interparticle friction is expected to play a dominant role \cite{DeGiuli16},
			and the physics of the system is governed by different effects.
			In that sense, $\mu_1$ in the law Eq.~(\ref{eqMuexp}) should be interpreted as an asymptotic
			value, that could be reached in the $\mathcal{I}\rightarrow0$ limit, rather than the true limit
			that $\mu$ reaches, which can be a priori modified in the friction dominated regime.

\section{Conclusion}

	All in all, in this work it is shown how the GITT formalism presented in \cite{Kranz20} can be used to describe
	the rheology of granular liquids, namely granular flows dense enough so that their physics is qualitatively different from
	that of gases, but dilute enough so that more subtle effects involving interparticle friction do not come into play.
	In addition to the shear stress, Eq.~(\ref{eqRheoA}) that has already been derived in \cite{Kranz20},
	here we have derived a generalized Green-Kubo relation for the pressure, Eq.~(\ref{eqRheoB}).
	The addition of the pressure to the set of equations is crucial insofar as it opens the possibility
	to compute quantities such as $\mu$, which are central to the understanding of the rheology of granular liquids.
	The addition of the pressure equation in Eq.~(\ref{eqRheo}) introduced a new source of precision loss, that resulted in the
	impossibility to test adequately the dilatancy law in the very dense regime.
	This pinpoints the need for more precise determination of the properties of dense hard sphere flows.
	More work is required in that direction.

	Nevertheless, the good agreement of the $\mu(\mathcal{I})$ predictions with experimental data assess that the above issue is
	not of paramount importance for the theoretical description of the properties of granular liquid flows.
	In particular, the proximity between the theoretical estimate and the values measured on glass beads strongly suggest
	that the whole $\mu(\mathcal{I})$ pertaining the physics of granular liquids stems from collective behaviors relevant at the
	macroscopic scale that largely wash out many of the microscopic
	characteristics of the packing such as its restitution coefficient, size distribution, driving mechanism, or internal friction.
	This is probably the reason why such a set of very simple constitutive laws,
	or why a simple theory of frictionless monodisperse hard spheres are able to capture the relevant physics of such flows.
	To that extent, our study thus represents a first step towards the design of a theory of granular liquids.

\section*{Acknowledgments}

	This work was funded by the Deutscher Akademischer Austauschdienst (DAAD) and the Deutsche Forschungsgemeinschaft (DFG),
	grant KR 486712.
	Matthias Sperl acknowledges discussions during "Granular Matter Across Scales", Lorentz Center, Leiden University,
	Leiden, The Netherlands, March 18-22, 2019.

\appendix

\section{Mode Coupling Equation}
\label{AMCT}

	MCT provides an evolution equation for the dynamics of the transient correlator.
	Its derivation is quite subtle in the granular case \cite{Kranz10,Kranz13,Kranz20}, we give here only the result:
	\begin{equation}
	\label{eqMCT}
		\begin{split}
			\ddot{\Phi}_q(t) &+ \nu_{q(t)}\dot{\Phi}_q(t)+q^2(t) C_{q(t)}^2 \Phi_q(t) \\
			+&q^2(t) C_{q(t)}^2 \int_0^td\tau\,m_q(t,\tau)\dot{\Phi}_q(\tau) =  0 \:.
		\end{split}
	\end{equation}
	The first three term represent the evolution of a quite general relaxation process, with a damping term given by \cite{Kranz13}:
	\begin{equation}
		\nu_q = \frac{1+\varepsilon}{3} \,\omega_c\,\Big[1+3j_0''(qd)\Big]\:,
	\end{equation}
	(where $d$ is the particle's diameter and $j_0$ is the zeroth-order spherical Bessel function), and a speed of sound~:
	\begin{equation}
		C_q^2 = \frac{T}{S_q}\left[\frac{1+\varepsilon}{2}+\frac{1-\varepsilon}{2}\,S_q\right]\:.
	\end{equation}
	The last term in Eq.~(\ref{eqMCT}) is a memory term that encodes non-markovian effects.
	It is defined by a memory kernel, that can be written as \cite{Kranz20}:
	\begin{equation}
	\label{eqmMCT}
		\begin{split}
			m_q(t,\tau) &= A_{q(t)}(\varepsilon)\frac{S_{q(t)}}{nq^2}\int\frac{d^3k}{(2\pi)^3} S_{k(\tau)}S_{p(\tau)} \\
			&\times\big[(\hat{\mathbf{q}}.\mathbf{k})nc_{k(t)} + (\hat{\mathbf{q}}.\mathbf{p})nc_{p(t)}\big]\\
			&\times\big[(\hat{\mathbf{q}}.\mathbf{k})nc_{k(\tau)} + (\hat{\mathbf{q}}.\mathbf{p})nc_{p(\tau)}\big] \\
			&\times\Phi_{k(\tau)}(t-\tau)\Phi_{p(\tau)}(t-\tau) \,,
		\end{split}
	\end{equation}
	where $n$ is the fluid's density, hats denote normalized vectors, $c_q$ denote the direct correlation function,
	and $A_q(\varepsilon)$ is a prefactor given by \cite{Kranz13}:
	\begin{equation}
	\label{eqA}
		A_q^{-1}(\varepsilon) = 1+\frac{1-\varepsilon}{1+\varepsilon}\,S_q\:.
	\end{equation}

\bibliography{MuI.bib}

\begin{thebibliography}{85}
\bibitem[{foo({\natexlab{a}})}]{foot2}
\bibinfo{note}{In their original works \cite{Fuchs02,Fuchs03, Fuchs09} Fuchs
  and Cates called Wi the "dressed Peclet number", thereby emphasizing the
  crucial role of the cage effect because of which $\omega_c$ is not the right
  scale to describe the relaxation of $\Phi_k(t)$. In that sense, fluctuations
  renormalize the time scale associated to structural relaxations at high
  densities.}
\bibitem[{foo({\natexlab{b}})}]{foot3}
\bibinfo{note}{Since $\sigma$ is defined from a zero-momentum limit of the
  vertices, the usual projectors are equal to the identity tensor.}
\bibitem[{foo({\natexlab{c}})}]{foot1}
\bibinfo{note}{In these expressions, the triplet structure factor
  $S^{(3)}(k,-k)$ has been replaced by the factorization ansatz, as done in
  standard calculations. Note that in this case the total circulating momentum
  is equal to zero, so that the triplet structure factor could have been
  expressed exactly. In this study we decided to keep the approximate
  expression to ensure consistency whith the MCT equations in which a non-zero
  momentum circulates and the factorization ansatz is required in the
  expression of the vertex.}
\bibitem[{foo({\natexlab{d}})}]{foot5}
\bibinfo{note}{In order to be able to display the evolution of the
  dimensionless pressure in the Bagnold regime, we used the Protocol H (keeping
  the driving power fixed) rather than the Protocol T (keeping $T$ fixed) used
  in the rest of the paper (see \cite{Kranz20} for details).}
\bibitem[{foo({\natexlab{e}})}]{foot4}
\bibinfo{note}{The collision frequency can be estimated from the packing
  fraction by using the Enskog expression $\omega_c = 24\varphi\chi
  d^{-1}\sqrt{T/\pi}$ \cite{Hansen06} where $\chi$ is the contact value of the
  pair correlation function. In this work, $\chi$ has been estimated from its
  expression in the P[4/5] ansatz, to be consistent with the equation of
  state.}
\end{thebibliography}

\end{document}